\documentstyle[12pt,epsfig]{article}
\begin{document}
\centerline{\Large\bf Time variation of the gravitational
coupling constant}
\vspace*{0.050truein}
\centerline{\Large\bf in decrumpling cosmology}
\vspace*{0.050truein}
\centerline{Forough Nasseri\footnote{Email: nasseri@fastmail.fm}}
\centerline{\it Institute for Astrophysics, P.O.Box 769, Neishabour, Iran}
\begin{center}
(\today)
\end{center}

\begin{abstract}
Within the framework of a model universe with time variable space
dimension (TVSD) model, known as decrumpling or TVSD model, we study the
time variation of the gravitational coupling constant. Using observational
bounds on the present time variation of the gravitational Newton's
constant in three-dimensional space we are able to obtain a constraint
on the time variation of the gravitational coupling constant.
As a result, the absolute value of the time variation of the
gravitational coupling constant must be less than
$\sim 10^{-11} {\rm yr}^{-1}$.\\
\end{abstract}
\vspace*{0.005truein}
\section{Introduction}
Usually in cosmological models based on higher dimensions the problem of
dimensionality of the gravitational coupling constant is not tackled on, being
tacitly assumed to be
\begin{equation}
\label{1}
\kappa=8 \pi G,
\end{equation}
where $G$ is the Newton's gravitational constant. This is of course a
relationship being derived in $(3+1)$-dimensional spacetime.
In this paper, we first review time variation of the gravitational
coupling constant in all dimensions 
and then study time variation of the gravitational coupling
constant in a model universe with time variable space dimensions (TVSD),
known as decrumpling or TVSD model.

The plan of this paper is as follows. In Section 2, we 
review the gravitational coupling constant in all dimensions.
In Section 3, we first 
review decrumpling or TVSD model and then obtain the time variation of
the gravitational coupling constant in this model.
Finally, we discuss our results and conclude in Section 4.
We will use a natural unit system that sets $k_B$, $c$, and $\hbar$
are equal to one, so that $\ell_P=M_P^{-1}=\sqrt{G}$.

\section{Gravitational coupling constant
in all dimensions}

Let us here review the gravitaional coupling constant in all dimensions
(see Ref. \cite{1}).
Take the metric in $(D+1)$-dimensional spacetime in the following
form 
\begin{equation}
\label{2}
ds^2= - dt^2 +a^2(t) d \Sigma_k^2,
\end{equation}
where $d \Sigma_k^2$ is the line element for a $D$-manifold of constant
curvature $k=-1, 0, +1$, corresponding to the closed, flat and hyperbolic
spacelike sections, respectively.
Using Eq.(\ref{2}), we obtain\footnote{It is worth mentioning that
in $D$-dimensional spaces the authors of Ref.\cite{2} have considered
$R_{00}= \nabla^2 \phi$. For this reason, the results presented in
Ref.\cite{2} are not in agreement with the results presented here
and Ref.\cite{1}.}
\begin{equation}
\label{3}
R_{00}= (D-2) \nabla^2_{D} \phi,
\end{equation}
where $\nabla_D$ is the $\nabla$ operator in $D$-dimensional spaces. In
$(3+1)$-dimensional spacetime, the Poisson equation is given by
$\nabla^2 \phi = 4 \pi G \rho$. Applying Gauss law for a $D$-dimensional
volume, we find the Poisson equation for arbitrary fixed dimension
\begin{equation}
\label{4}
\nabla^2_D \phi = S^{[D]} G_{(D+1)} \rho,
\end{equation}
where $G_{(D+1)}$ is the $(D+1)$-dimensional Newton's constant and
$S^{[D]}$ is the surface area of a unit sphere in $D$-dimensional
spaces
\begin{equation}
\label{5}
S^{[D]} = \frac{2 \pi^{D/2}}{\Gamma \left( \frac{D}{2} \right) }.
\end{equation}
On the other hand from Eq.(\ref{2}) we get
\begin{equation}
\label{6}
R_{00} = \left( \frac{D-2}{D-1} \right) \kappa_{(D+1)} \rho.
\end{equation}
Using Eqs.(\ref{3}-\ref{6}), we are led to the gravitational coupling
constant in $(D+1)$-dimensional spacetime\footnote{For example,
in the cases $(3+1)$, $(4+1)$ and $(5+1)$-dimensional spacetime
we have $\kappa_{(3+1)}=8 \pi G_{(3+1)}$ (i.e. $\kappa = 8 \pi G$),
$\kappa_{(4+1)} = 6 \pi^2 G_{(4+1)}$ and $\kappa_{(5+1)}=
\frac{32 \pi^2}{3} G_{(5+1)}$, respectively.}
\begin{equation}
\label{7}
\kappa_{(D+1)} = (D-1) S^{[D]} G_{(D+1)}=
\frac{2 (D-1) \pi^{D/2} G_{(D+1)}}{\Gamma \left( \frac{D}{2} \right)}.
\end{equation}
Let us now obtain a relationship between the Newton's constant in
$(D+1)$-dimensional spacetime and in the $(3+1)$-dimensional spacetime.
Using the force laws in $(D+1)$- and $(3+1)$- dimensional spacetime,
which are defined by
\begin{eqnarray}
\label{8}
F_{(D+1)} (r) &=& G_{(D+1)} \frac{m_1 m_2}{r^{D-1}},\\
\label{9}
F_{(3+1)} (r) &=& G_{(3+1)} \frac{m_1 m_2}{r^2},
\end{eqnarray}
and the $(D+1)$-dimensional Gauss law one can derive (see Ref.\cite{3})
\begin{equation}
\label{10}
G_{(3+1)}= \frac{S^{[D]}}{4 \pi} \frac{G_{(D+1)}}{V^{[D-3]}},
\end{equation}
where $V^{[D-3]}$ is the volume of $(D-3)$ extra spatial dimensions.
Now, using Eqs.(\ref{7}) and (\ref{10}) we are led to
\begin{equation}
\label{11}
\kappa_{(D+1)} = 4 \pi (D-1) G_{(3+1)} V^{[D-3]}.
\end{equation}
Our approach here to obtain the gravitational coupling
constant in all dimensions is a model-independent approach
and may be used for cosmological models in higher dimensions.

\section{Time variation of gravitational coupling constant in the model}

Let us briefly review decrumpling or TVSD model (for more details
see Refs.\cite{4}-\cite{11}).
Assume the universe consists of a fixed number ${\bar N}$ of universal
cells having a characteristic length $\delta$ in each of their
dimensions. The volume of the universe at the time $t$ depends
on the configuration of the cells. It is easily seen that
\cite{10}
\begin{equation}
\label{12}
{\rm vol}_{D_t}({\rm cell})={\rm vol}_{D_0}({\rm cell})\delta^{D_t-D_0},
\end{equation}
where the $t$ subscript in $D_t$ means $D$ is to be as a function
of time.
Interpreting the radius of the universe, $a$, as the radius of
gyration of a crumpled ``universal surface'',
the volume of space can be written \cite{10}
\begin{eqnarray}
\label{13}
a^{D_t}&=&{\bar N} {\rm vol}_{D_t}({\rm cell})\nonumber\\
   &=&{\bar N} {\rm vol}_{D_0}({\rm cell}) \delta^{D_t-D_0}\nonumber\\
   &=&{a_0}^{D_0} \delta^{D_t-D_0}
\end{eqnarray}
or
\begin{equation}
\label{14}
\left( \frac{a}{\delta} \right)^{D_t}=
\left( \frac{a_0}{\delta} \right)^{D_0} = e^C,
\end{equation}
where $C$ is a universal positive constant. Its value has a strong
influence on the dynamics of spacetime, for example on the dimension
of space, say, at the Planck time. Hence, it has physical and cosmological
consequences and may be determined by observations. The zero subscript in any
quantity, e.g. in $a_0$ and $D_0$, denotes its present values.
We coin the above relation as a``dimensional constraint" which relates
the ``scale factor" of the model universe to the space dimension.
In our formulation, we consider the comoving length of the Hubble radius
at present time to be equal to one. So the interpretation of the scale
factor as a physical length is valid.
The dimensional constraint can be written in this form
\begin{equation}
\label{15}
\frac{1}{D_t}=\frac{1}{C}\ln \left( \frac{a}{a_0} \right) + \frac{1}{D_0}.
\end{equation}

It is seen that by expansion of the universe, the space
dimension decreases. 
Time derivative of Eqs.(\ref{14}) or (\ref{15}) leads to
\begin{equation}
\label{16}
{\dot{D}}_t=-\frac{D_t^2 \dot{a}}{Ca}.
\end{equation}
It can be easily shown that the case of constant space dimension
corresponds to when $C$ tends to infinity. In other words,
$C$ depends on the number of fundamental cells. For $C \to +\infty$,
the number of cells tends to infinity and $\delta\to 0$.
In this limit, the dependence between the space dimensions and
the radius of the universe is removed, and consequently we
have a constant space dimension.

We define $D_P$ as the space dimension of the universe when the
scale factor is equal to the Planck length $\ell_P$.
Taking $D_0=3$ and the scale of the universe today to be the present
value of the Hubble radius $H_0^{-1}$ and the space dimension at the
Planck length to be $4, 10,$ or $25$, from Kaluza-Klein and superstring
theories, we can obtain from Eqs. (\ref{15}) and (\ref{16})
the corresponding value of $C$ and $\delta$
\begin{eqnarray}
\label{17}
\frac{1}{D_P}&=& \frac{1}{C} \ln \bigg( \frac{\ell_P}{a_0}\bigg) +
\frac{1}{D_0}= \frac{1}{C} \ln \bigg( \frac{\ell_P}{H_0^{-1}}\bigg) +
\frac{1}{3},\\
\label{18}
\delta&=&a_0 e^{-C/D_0}=H_0^{-1} e^{-C/3}.
\end{eqnarray}
In Table 1, values of $C$, $\delta$ and also
${\dot D}_t|_0$ for some interesting values of $D_P$ are
given. These values are calculated by
assuming $D_0=3$ and
$H_0^{-1}=3000 {h_0}^{-1} {\rm Mpc} = 9.2503 \times 10^{27} {h_0}^{-1} 
{\rm cm}$, where $h_0=0.68 \pm 0.15$.
Since the value of $C$ and $\delta$ are not very
sensitive to $h_0$ we take $h_0=1$.
\begin{table}
\caption{Values of $C$ and $\delta$ for some values of
$D_P$ \cite{4}-\cite{10}. Time variation of space dimension today
has also been calculated in terms of 
yr$^{-1}$.}
\begin{tabular}{cccc} \\ \hline\hline 
$D_P$ & $C$   & $\delta$ (cm)   & $\dot D_t|_0$ (yr$^{-1}$) \\ \hline\hline
$3$           & $ +\infty$         &  $0$           & $0$ \\ \hline
$4$           & $1678.797$         &  $8.6158 \times 10^{-216}$  & $ -5.4827 \times 10^{-13} h_0$  \\ \hline
$10$          & $599.571$          &  $1.4771 \times 10^{-59}$  & $ -1.5352 \times 10^{-12} h_0$  \\ \hline
$25$          & $476.931$          &  $8.3810 \times 10^{-42}$  & $-1.9299 \times 10^{-12} h_0$ \\ \hline
$+\infty$     & $419.699$          &  $\ell_P$  & $ -2.1931 \times 10^{-12}h_0$ \\ \hline\hline
\end{tabular}
\end{table}

Let us define the action of the model for the special
Friedmann-Robertson-Walker (FRW) metric in an arbitrary fixed space
dimension $D$, and then try to generalize it to variable dimension $D_t$.
Now, take the metric in constant $D+1$ spacetime dimensions in the
following form
\begin{equation}
\label{19}
ds^2 = -N^2(t)dt^2+a^2(t)d\Sigma_k^2,
\end{equation}
where $N(t)$ denotes the lapse function and $d\Sigma_k^2$ is the line
element for a D-manifold of constant curvature $k = + 1, 0, - 1$. The
Ricci scalar is given by
\begin{equation}
\label{20}
R=\frac{D}{N^2}\left\{\frac{2\ddot a}{a}+(D-1)\left[\left(\frac{\dot a}{a}
\right)^2 + \frac{N^2k}{a^2}\right]-\frac{2\dot a\dot N}{aN}\right\}.
\end{equation}
Substituting from Eq.(\ref{20}) in the Einstein-Hilbert action for
pure gravity,
\begin{equation}
\label{21}
S_G = \frac{1}{2\kappa} \int d^{(1+D)} x \sqrt{-g}R,
\end{equation}
and using the Hawking-Ellis action of a perfect fluid
for the model universe with variable space dimension the following
Lagrangian has been obtained for decrumpling or TVSD model
(see Ref.\cite{10})
\begin{equation}
\label{22}
L_I := -\frac{V_{D_t}}{2 \kappa N} \left( \frac{a}{a_0} \right)
^{D_t} D_t(D_t-1)
\left[ \left( \frac{\dot a}{a} \right )^2 -\frac{N^2 k}{a^2} \right ]
- \rho N V_{D_t} \left( \frac{a}{a_0} \right )^{D_t},
\end{equation}
where $\kappa=8 \pi {M_P}^{-2}=8 \pi G$, $\rho$ the energy density,
and $V_{D_t}$ the volume of the space-like sections

\begin{eqnarray}
\label{23}
V_{D_t}&=&\frac{2 \pi^{(D_t+1)/2}}{\Gamma[(D_t+1)/2]},\;\;\mbox{closed Universe, $k=+1$,}\\
\label{24}
V_{D_t}&=&\frac{\pi^{(D_t/2)}}{\Gamma(D_t/2+1)}{\chi_c}^{D_t},\;\;\mbox{flat Universe, $k=0$,}\\
\label{25}
V_{D_t}&=&\frac{2\pi^{(D_t/2)}}{\Gamma(D_t/2)}f(\chi_c),\;\;\mbox{open Universe, $k=-1$.}
\end{eqnarray}

Here $\chi_C$ is a cut-off and $f(\chi_c)$ is a function thereof
(see Ref. \cite{10}).

In the limit of constant space dimensions, or $D_t=D_0$,
$L_I$ approaches to the Einstein-Hilbert Lagrangian
which is
\begin{equation}
\label{26}
L_{I}^0 := - \frac{V_{D_0}}{2 \kappa_0 N}
\left( \frac{a}{a_0} \right)^{D_0} D_0(D_0-1)
\left[ \left( \frac{\dot{a}}{a} \right)^2 - \frac{N^2 k}{a^2} \right ]
- \rho N V_{D_0} \left( \frac{a}{a_0} \right )^{D_0},
\end{equation}
where $\kappa_0=8\pi G_0$ and the zero subscript in $G_0$ denotes its
present value. So, Lagrangian $L_I$ cannot abandon Einstein's gravity.
Varying the Lagrangian $L_I$ with respect to $N$ and $a$, we find the
following equations of motion in the gauge $N=1$, respectively (see
Ref.\cite{10})
\begin{eqnarray}
\label{27}
&&\left( \frac{\dot a}{a} \right)^2 +\frac{k}{a^2} =
\frac{2 \kappa \rho}{D_t(D_t-1)},\\
\label{28}
&&(D_t-1) \bigg\{ \frac{\ddot{a}}{a} + \left[ \left( \frac{\dot a}{a}
\right)^2
+\frac {k}{a^2} \right] \bigg( -\frac{{D_t}^2}{2C} \frac{d \ln V_{D_t}}{d{D_t}}
-1-\frac{D_t(2D_t-1)}{2C(D_t-1)} 
+\frac{{D_t}^2}{2D_0} \bigg) \bigg\} \nonumber\\
&&+ \kappa p \bigg( -\frac{d \ln V_{D_t}}{d{D_t}} \frac{D_t}{C} 
-\frac{D_t}{C} \ln \frac{a}{a_0} +1 \bigg) =0.
\end{eqnarray}
Using (\ref{16}) and (\ref{27}), the evolution equation of the space
dimension can be obtained by
\begin{equation}
\label{29}
{{\dot{D}}_t}^2= \frac{D_t^4}{C^2} \left[ \frac{2 \kappa \rho}{D_t(D_t-1)}
-k {\delta}^{-2} e^{-2C/{D_t}} \right].
\end{equation}
The continuity equation of decrumpling or TVSD model
can be obtained by (\ref{27}) and (\ref{28})
\begin{equation}
\label{30}
\frac{d}{dt} \left[ \rho \left( \frac{a}{a_0} \right)^{D_t} V_{D_t} \right]
+ p \frac{d}{dt} \left[ \left( \frac{a}{a_0} \right )^{D_t} V_{D_t} \right] =0.
\end{equation}

In TVSD model, we can rewrite Eq.(\ref{10}) in this form
\begin{equation}
\label{31}
G_{(3+1)}= \frac{S^{[D_t]}}{4 \pi} \frac{G_{(D_t+1)}}{V^{[D_t-3]}},
\end{equation}
where $V^{[D_t-3]}$ is the volume of $(D_t-3)$ extra spatial dimensions.
One general feature of extra-dimensional theories, such as
Kaluza-Klein and string theories, is that the ``true'' constants
of nature are defined in the full higher dimensional theory
so that the effective $4$-dimensional constants depends,
among other things, on the structure and size of the extra-dimensions.
Any evolution of these sizes either in time or space, would lead
to a spacetime dependence of the effective $4$-dimensional constants,
see Ref. \cite{12}.
So $G_{(D_t+1)}$ is a ``true'' constant and
has variable dimension [length]$^{D_t-1}$. Therefore, we have
${\dot G}_{(D_t+1)}=0$.

Using Eq.(\ref{31}), the time variation of the gravitational Newton's
constant in $(3+1)$-dimensional spacetime and in TVSD model is given by
\begin{equation}
\label{32}
\frac{\dot G_{(3+1)}}{G_{(3+1)}}=
\frac{{\dot S}^{[D_t]}}{S^{[D_t]}}-
\frac{{\dot V}^{[D_t-3]}}{V^{[D_t-3]}}.
\end{equation}
On the other hand, in TVSD model the surface area of a unit sphere in
$D_t$-dimensional spaces is given by (see Eq.(\ref{5}))
\begin{equation}
\label{33}
S^{[D_t]} = \frac{2 \pi^{D_t/2}}{\Gamma \left( \frac{D_t}{2} \right) }.
\end{equation}
Time derivative of this equation leads to
\begin{equation}
\label{34}
\frac{{\dot S}^{[D_t]}}{S^{[D_t]}}=
\frac{{\dot D}_t}{2} \left[ \ln \pi - \psi \left( \frac{D_t}{2} \right) \right],
\end{equation}
where Euler's psi function $\psi$ is the logarithmic derivative of the
gamma function $\psi(x)\equiv \Gamma'(x)/\Gamma (x)$.

In TVSD model, the volume of $(D_t-3)$ extra spatial dimensions
is given by\footnote{In Ref.\cite{4}, we considered
$V^{[D_t-3]} \simeq a^{D_t-3}$. This makes
some problems in calculations of Ref.\cite{4}.
For example, Eq.(\ref{37}) in Ref.\cite{4} must be corrected to
Eq.(\ref{37}) in this paper.}
(see Eq.(\ref{22}))
\begin{equation}
\label{35}
V^{[D_t-3]} = \left( \frac{a}{a_0} \right)^{D_t-3},
\end{equation}
where zero subscript in $a_0$ denotes the present value
of the scale factor. Therefore, we have
\begin{equation}
\label{36}
\frac{{\dot V}^{[D_t-3]}}{V^{[D_t-3]}}=
{\dot D}_t \ln \left( \frac{a}{a_0} \right)
+ (D_t-3) \frac{\dot a}{a}.
\end{equation}
Using Eqs.(\ref{32}), (\ref{34}) and (\ref{36}) we obtain
\begin{equation}
\label{37}
\frac{\dot G_{(3+1)}}{G_{(3+1)}}=
\frac{{\dot D}_t}{2} \left[ \ln \pi - \psi \left( \frac{D_t}{2} \right)
\right]- {\dot D}_t \ln \left( \frac{a}{a_0} \right)
- (D_t-3) \frac{\dot a}{a}.
\end{equation}
Let us obtain the time variation of the gravitational coupling
constant in decrumpling or TVSD model.
From Eq.(\ref{11}), it can be easily obtained the gravitational coulping
constant in TVSD model
\begin{equation}
\label{38}
\kappa_{(D_t+1)} = 4 \pi (D_t-1) G_{(3+1)} V^{[D_t-3]}.
\end{equation}
Time derivative of Eq.(\ref{38})
leads to
\begin{equation}
\label{39}
\frac{{\dot \kappa}_{(D_t+1)}}{\kappa_{(D_t+1)}}=
\frac{{\dot D}_t}{(D_t-1)}+
\frac{{\dot G}_{(3+1)}}{G_{(3+1)}}+
\frac{{\dot V}^{[D_t-3]}}{V^{[D_t-3]}}.
\end{equation}
From Eqs.(\ref{32}) and (\ref{39}), we obtain
\begin{equation}
\label{40}
\frac{{\dot \kappa}_{(D_t+1)}}{\kappa_{(D_t+1)}}=
\frac{{\dot D}_t}{(D_t -1)} +
\frac{{\dot S}^{[D_t]}}{S^{[D_t]}}.
\end{equation}
Using Eqs.(\ref{34}) and (\ref{40}) we obtain
\begin{equation}
\label{41}
\frac{{\dot \kappa}_{(D_t+1)}}{\kappa_{(D_t+1)}} =
{{\dot D}_t}
\left[ \frac{1}{D_t-1} + \frac{\ln \pi}{2}
- \frac{1}{2} \psi \left( \frac{D_t}{2} \right) \right].
\end{equation}
Let us now use Eqs.(\ref{37}) and (\ref{41}) at the present time.
Taking $D_t|_0=3$, $a=a_0$ and using
\begin{equation}
\label{42}
\psi \left( \frac{3}{2} \right)=2 - \gamma -2 \ln(2),
\end{equation}
where $\gamma$ is Euler's constant and approximately $0.5772156649$,
we obtain from Eqs.(\ref{37}) and (\ref{41}) respectively
\begin{eqnarray}
\label{43}
\frac{\dot G_{(3+1)}}{G_{(3+1)}} & \simeq &
0.554\,{{\dot D}_t},\\
\label{44}
\frac{{\dot \kappa}_{(D_t+1)}}{\kappa_{(D_t+1)}} & \simeq &
1.054\,{{\dot D}_t}.
\end{eqnarray}
According to Ref.\cite{12}, the absolute value of the time variation
of the Newton's constant in three-space dimension today has an upper
limit
\begin{equation}
\label{45}
\bigg| \frac{\dot G_{(3+1)}}{G_{(3+1)}} \bigg| \Bigg|_0 <
9 \times 10^{-12} {\rm yr}^{-1}.
\end{equation}
Using Eqs.(\ref{43}) and (\ref{44}) one gets
\begin{equation}
\label{46}
\bigg| {\dot D}_t \bigg| \Bigg|_0 < 1.6 \times 10^{-11} {\rm yr}^{-1}.
\end{equation}
Now, from Eqs.(\ref{44}) and (\ref{46}) we obtain
\begin{equation}
\label{47}
\bigg| \frac{{\dot \kappa}_{(D_t+1)}}{\kappa_{(D_t+1)}} \bigg| \Bigg|_0
< 1.7 \times 10^{-11} {\rm yr}^{-1}.
\end{equation}
\section{Conclusions}
In this paper, we study the time variation of the gravitational
coupling constant in TVSD or decrumpling model.
Using observational bounds on the time variation of the Newton's
constant in three-space dimension we obtain a constraint on the
absolute value of the
time variation of the spatial dimension, see Eq.(\ref{46}),
and then a constraint on the absolute value of the time variation
of the gravitational coupling constant in the model.
It is worth mentioning that in Ref.\cite{11},
by using the observational bounds
on the time variation of the fine structure constant,
we have obtained
a constraint on the absolute value of the time variation of the
spatial dimension which is
$|{\dot D}_t||_0 < 10^{-15} {\rm yr}^{-1}$.
Comparing our result in this paper, i.e. Eq.(\ref{46}), with the
result presented in Ref.\cite{11}, one can conclude the absolute value
of the time variation of the spatial dimension must be less than
$10^{-15} {\rm yr}^{-1}$.\\
\noindent
{\bf Acknowledgments:} F.N. thanks Hurieh Husseinian
and Ali Akbar Nasseri for noble helps.

\end{document}